\documentclass[12pt]{article}

\catcode`\@=11
\@addtoreset{equation}{section}

\global\arraycolsep=2pt

\usepackage{mathrsfs,amsbsy,amssymb,latexsym,amsfonts,amsmath,cite}
\usepackage{graphicx,color}

\newcommand{\dd}{{\rm d}}
\newcommand{\gym}{g_{\text{YM}}}

\newcommand{\ads}{\text{AdS} }
\newcommand{\tf}{T_{\rm F} }

\newcommand{\rc}{r_{\rm c} }

\newcommand{\rt}{r_{\rm t} }

\allowdisplaybreaks

\begin{document}
\begin{flushright}
\parbox{4.2cm}
{KUNS-2464}
\end{flushright}

\vspace*{2cm}

\begin{center}
{\Large \bf Universal aspects of holographic Schwinger effect 
\\ in general backgrounds}
\vspace*{2cm}\\
{\large Yoshiki Sato\footnote{E-mail:~yoshiki@gauge.scphys.kyoto-u.ac.jp} 
and 
Kentaroh Yoshida\footnote{E-mail:~kyoshida@gauge.scphys.kyoto-u.ac.jp} 
}
\end{center}

\vspace*{1cm}
\begin{center}
{\it Department of Physics, Kyoto University \\ 
Kyoto 606-8502, Japan} 
\end{center}

\vspace{1cm}

\begin{abstract}
We consider universal aspects of a holographic Schwinger effect 
in general backgrounds with an external homogeneous electric field. 
The argument is based on the potential analysis 
developed in our previous work. Under some conditions, 
there always exists a critical electric field, above which the potential barrier vanishes  
and the system becomes unstable catastrophically. The critical value agrees with the one obtained 
from the Dirac-Born-Infeld action. For general confining backgrounds, we show that the 
Schwinger effect does not occur when the electric field is weaker 
than the confining string tension. 
\end{abstract}

\thispagestyle{empty}
\setcounter{page}{0}
\setcounter{footnote}{0}

\newpage

\section{Introduction}

The Schwinger effect \cite{Schwinger} is the pair production of 
electron and positron in quantum electro dynamics (QED) 
in the presence of a strong external electric field. 
This is known as a non-perturbative phenomenon.  
The production rate is computed under weak-field 
and weak-coupling conditions \cite{Schwinger}, and then it is computed only 
by a weak-field approximation \cite{AAM} (For the monopole production, see \cite{AM}). 
The pair production is not restricted to QED, but ubiquitous 
in quantum field theories coupled to an abelian gauge field. 

\medskip 

Pairs of virtual particle and anti-particle momentarily created in the vacuum must 
gain energy more than the total static mass so as to be real particles. 
This process is described as a tunneling phenomenon 
with the potential, 
\[
V(x)=2m-Ex-\frac{\alpha _{\rm s}}{x}\,.
\] 
Here the first term is the total static mass and then the second is the energy coming from 
an external electric field $E$\,. Finally the third is the Coulomb potential 
with the fine-structure constant $\alpha _{\rm s}$\,. 
The potential barrier decreases gradually as the electric field becomes stronger, 
and at last it vanishes at a certain value of the electric field $E=E_{\rm c}$\,.  
This value is called the critical electric field, above which there is no potential barrier 
and the system becomes unstable catastrophically.

\medskip 

A motive is to understand the critical electric field in the context of string theory \cite{max1,max2}.   
The AdS/CFT correspondence \cite{M,GKP,W} provides a nice laboratory \cite{GSS}. 
Semenoff and Zarembo recently proposed a holographic setup to study 
the Schwinger effect \cite{SZ}. In this setup, the system is the $SU(N+1)$ $\mathcal{N}=4$ 
super Yang-Mills theory and the gauge group 
$SU(N+1)$ is broken to $SU(N) \times U(1)$ by the Higgs mechanism. 
A probe D3-brane is put at an intermediate position in the bulk AdS space 
and the mass of the fundamental particles (called ``quarks'') is finite. 
At  large $N$ and large 't~Hooft coupling $\lambda \equiv \gym ^2N$\,,   
the production rate of quarks with mass $m$ 
is estimated as \cite{SZ}\footnote{The prefactor has not been obtained yet because of the difficulty 
to evaluate quantum fluctuations around a circular Wilson loop. For attempts along this direction, 
see \cite{DGT,SaY,AmMa,KM}.}
\begin{equation}
\Gamma \sim \exp \left[ -\frac{\sqrt{\lambda}}{2}
\left( \sqrt{\frac{E_{\text{c}} } {E}}-\sqrt{\frac{E } {E_{\text{c}}}} \, \right)^2\right]
 \,, \qquad E_{\text{c}}=\frac{2\pi m^2}{\sqrt{\lambda}}\,. 
\end{equation}
Here $E_{\rm c}$ is the critical electric field and agrees with the result obtained 
from the Dirac-Born-Infeld (DBI) action. For generalizations of this setup to the case with magnetic fields 
and the finite temperature case, see \cite{BKR,SY}. 

\medskip 

We have revisited the agreement of the critical electric field 
from the viewpoint of the potential analysis \cite{SY2}\footnote{The potential at a specific value of the electric field 
is computed in \cite{KL}.}. 
The probe D3-brane is put at an intermediate position between the horizon and the boundary 
in the AdS space, as in \cite{SZ}. Then the quark anti-quark potential is computed 
from the vacuum expectation value (VEV) of a rectangular Wilson loop 
on the probe D3-brane by slightly generalizing the methods in \cite{Wilson1,Wilson2}. 
With this quark anti-quark potential, we have shown that the critical electric field agrees with 
$E_{\rm c}=2\pi m^2/\sqrt{\lambda}$\,. 

\medskip 

This potential analysis is quite a powerful tool because the computation of  
the VEV of a rectangular Wilson loop is much easier than that of a circular Wilson loop \cite{BCFM,DGO}. 
Hence the potential analysis is applicable to a wide class of systems. A good example is 
AdS solitons \cite{HM} which lead to confinement. 
For the confining backgrounds, we have shown that 
there is another critical value of the electric field \cite{SY3}. 
When the electric field is smaller than this value, the Schwinger effect does not occur. 

\medskip 

It is quite interesting to consider general arguments for the critical electric fields. 
In this paper we will carry out the potential analysis for general backgrounds and 
discuss two kinds of the critical electric field. The first is the critical electric field for the catastrophic decay. 
It universally exists under some conditions for the backgrounds and agrees with the DBI action. 
This is a generalization of \cite{SZ} to a certain class of holographic backgrounds. 
The second is the one, below which the Schwinger effect does not occur, 
in general confining backgrounds. 
The sufficient conditions for confining backgrounds are elaborated in \cite{Son1} 
(For reviews see \cite{Son2,Son3}). 
Under the conditions, the critical electric field universally exists. 

\medskip 

The organization of this paper is as follows. In section 2, we consider the critical electric field for 
the catastrophic decay in general backgrounds under some conditions. 
In section 3, we first introduce the theorem to specify confining backgrounds. 
For concreteness, two examples are explained. 
Then we discuss the critical electric field below which the Schwinger effect does not occur. 
Section 4 is devoted to conclusion and discussion.

\section{The catastrophic decay in general backgrounds}

We will consider the critical electric field concerning with 
the catastrophic vacuum decay in general supergravity backgrounds 
which are constructed basically from D$p$-branes and admit the holographic interpretation. 

\subsection{Setup}

Let us first introduce the setup for our argument. 
We are concerned with type IIB and IIA supergravity solutions 
obtained by taking the near-horizon limit of D$p$-brane backgrounds 
and their generalizations, for which a holographic interpretation is possible. 

\medskip 

We assume the diagonal metric for the D$p$-brane world-volume with time $t$ 
and spatial directions $x^i~(i=1,\ldots,p)$\,. The existence of the radial direction 
$r~(\rt \leq r < \infty)$ is also supposed to perform the holographic analysis, 
where $\rt$ is the position of the horizon or the infrared boundary for AdS solitons. 
For simplicity, we suppose that each component of the metric depends only on $r$\,.  
Thus the metric of the background in ten dimensions is basically given by 
\begin{equation}
\dd s^2=-G_{00}(r)\dd t^2+ \sum_{i=1}^pG_{ii}(r)(\dd x^i)^2+G_{rr}(r)\dd r^2+ \sum_{m,n=1}^{8-p}G_{mn}(r)\dd x^m \dd x^n\,, 
\label{metric}
\end{equation}
where $G_{mn}~(m,n=1,\ldots ,8-p)$ is the metric of the internal space in $8-p$ dimensions. 

\medskip 

The location $\rt$ is basically fixed as a zero point of $G_{00}(r)$ in the usual brane setup. 
For AdS soliton cases $\rt$ is given as a zero point of $G_{zz}(r)$\,, 
where $z$ is the compactified direction.

\subsection{The potential from the stringy computation}

Let us next consider the classical solution of the string world-sheet 
ending on a probe D-brane located at $r=r_0$\,. 
We will work in the Euclidean signature below. 

\medskip 

The Nambu-Goto (NG) action is given by
\begin{align}
S_{\rm NG}&=\tf \int \! \dd \tau \int \! \dd \sigma \, \mathcal{L} \,,\\
&\mathcal{L}=\sqrt{\det G_{ab}}\,, \qquad G_{ab}\equiv \frac{\partial x^{\mu}}{\partial \sigma ^a}
\frac{\partial x^{\nu}}{\partial \sigma ^b}\, g_{\mu \nu} 
\end{align}
where $\sigma ^a=(\tau ,\sigma)$ are the string world-sheet coordinates.

\medskip 

We are concerned with the potential between a quark and an anti-quark separated 
in the $x$-direction. For this purpose, it is convenient to take the static gauge, 
\begin{equation}
t = \tau\,, \qquad x=\sigma\,, 
\end{equation}
where $x$ is one of the non-compact spatial directions along the D-brane world-volume. 
The radial direction of the classical solution is supposed to depend only on $\sigma$\,, 
\begin{equation}
r=r(\sigma )\,.
\end{equation}
Under this ansatz, the analysis is drastically simplified. 

\medskip 

Using the spacetime metric \eqref{metric}\,, the Lagrangian is given by 
\begin{align}
{\cal L}&=\sqrt{f(r)^2+g(r)^2(\partial _{\sigma} r)^2} \,, \label{lag} 
\end{align}
where we have introduced functions $f(r)$ and $g(r)$ defined as, respectively, 
\begin{align}
&f(r) \equiv \sqrt{G_{00}(r)G_{xx}(r)} \,, \qquad g(r) \equiv \sqrt{G_{00}(r)G_{rr}(r)}\,. 
\end{align}
Here $f(r)$ and $g(r)$ are assumed to be real, smooth and non-negative 
so that the Lagrangian is well defined. 
For a holographic interpretation, we suppose a further condition that $f(r)$ is a monotonically increasing function. 
This condition indicates that the string solution should extend towards the horizon, as we will see later.

\medskip 

\begin{figure}[tbp]
\begin{center}
\includegraphics[scale=.5]{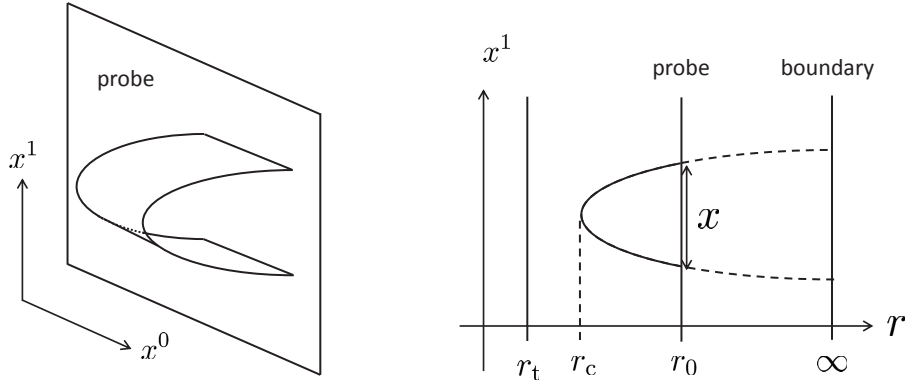}
\caption{\footnotesize The configuration of the string world-sheet. \label{AdS-soliton:fig}}
\end{center}
\end{figure}

Now that the resulting Lagrangian (\ref{lag}) depends only on $\sigma$\,, the Hamiltonian 
is a conserved quantity with respect to $\sigma$\,. Hence the following relation is obtained,  
\begin{equation}
{\cal L}-(\partial _\sigma r)\frac{ \partial {\cal L}}{ \partial (\partial _\sigma r)}
=\frac{f^2}{\sqrt{f^2+g^2(\partial _\sigma r)^2}}=\mbox{constant}\,. \label{const}
\end{equation}
By imposing the boundary condition,  
\begin{equation}
r=\rc \,, \quad \partial _\sigma r=0\quad (\sigma =0)\,,
\end{equation}
the constant term in (\ref{const}) can be fixed as 
\begin{equation} 
\frac{f^2}{\sqrt{f^2+g^2(\partial _\sigma r)^2}}=f(\rc) 
\end{equation}
and the differential equation is derived, 
\begin{equation}
\frac{\dd r}{\dd \sigma}=\pm \frac{f(r)}{g(r)}\frac{\sqrt{f(r)^2-f(\rc)^2}}{f(\rc)}\,. 
\end{equation}
Then the distance $x$ between a quark and an anti-quark is given by
\begin{equation}
x=2\int_{\rc} ^{r_0}\! \dd r \, \frac{g(r)}{f(r)}\frac{f(\rc)}{\sqrt{f(r)^2-f(\rc)^2}}\,, \label{x}
\end{equation}
where $r_0$ is the position of a probe D-brane. 
Assume that the above integral is finite. 
For the geometric relation of the parameters, see Figure \ref{AdS-soliton:fig}.

\medskip 

In the usual case, the probe D-brane is located near the boundary, namely $r_0 \to \infty$\,.  
Then the mass of ``quark'' is infinitely heavy.  
However, to discuss the pair production in a more realistic setup, 
the mass should be kept finite and hence the probe D-brane should be put 
at an intermediate position in the bulk space \cite{SZ}. 
Now that $r_0$ takes a finite value, the quark mass is given by 
\[
m=\tf\int_{\rt}^{r_0} \! \dd r \, g(r)\,.
\]
This describes the energy of a string stretched between the horizon and the probe D-brane. 

\medskip 

On the other hand, for the backgrounds that admit a holographic interpretation, 
the probe D-brane should be able to be sent to the boundary, in principle. 
So we assume that the distance $x$ in (\ref{x}) should be finite 
as a consistency of the background when the $r_0\to \infty$ limit is taken. 
A sufficient condition for this is given by \cite{Son1}
\begin{equation}
\int_{\tilde{r}}^{\infty}\! \dd r \, \frac{g(r)}{f(r)^2}<\infty \,, \label{finite}
\end{equation}
where $\tilde{r}$ is a certain constant satisfying $\tilde{r} > \rc$\,. 
It is easy to show the following inequalities,    
\begin{eqnarray}
1 < \frac{1}{\sqrt{1-f(\rc)^2/f(r)^2}} <  \frac{1}{\sqrt{1-f(\rc)^2/f(\tilde{r})^2}} \qquad \mbox{for}~~~~r > \tilde{r}\,.
\end{eqnarray}
Then, by multiplying $g(r)/f(r)^2$ and integrating from $\tilde{r}$ to $\infty$\,, the following integral
\begin{eqnarray}
\int_{\tilde{r}}^{\infty}\! \dd r \, \frac{g(r)}{f(r)^2}\frac{1}{\sqrt{1-f(\rc)^2/f(r)^2}} \label{f-int}
\end{eqnarray}
is shown to be finite due to the condition (\ref{finite})\,. 
With (\ref{f-int})\,, one can see that $x$ is finite when $r_0$ is sent to $\infty$\,. 
In other words, the condition (\ref{finite}) ensures the finiteness of the distance $x$ for an arbitrary mass, 
so that the quark and anti-quark potential can definitely be defined. Although it seems unnecessary to take care 
of heavy mass in studying the Schwinger effect, the pair creation may occur even for heavy particles in principle 
and we suppose the condition (\ref{finite}) for completeness.

\medskip 

According to the conventional AdS/CFT dictionary, 
the area of the string world-sheet ending on a rectangular Wilson loop leads to 
the sum of the quark anti-quark potential energy (PE) 
and the static energy (SE). This is also the case when the probe D-brane is located at an intermediate 
position $r=r_0$\,, rather than near the boundary. The resulting potential is given by   
\begin{equation}
V_{\rm PE +SE}= \tf \int \! \dd \sigma \,  \mathcal{L}
=2 \tf\int _{\rc}^{r_0} \! \dd r \, \frac{g(r)f(r)}{\sqrt{f(r)^2-f(\rc)^2}} \,.
\label{potential}
\end{equation}
 
\subsection{The critical electric field from the potential analysis \label{cri:sec}}

We shall evaluate the critical electric field by turning on a constant NS-NS two form 
$B_{tx} = 2\pi \alpha' E$\,. 
By rewriting \eqref{potential}\,, the total potential $V_{\rm tot}$ is given by 
\begin{align}
V_{\rm tot} & \equiv V_{\rm PE+SE} - E x \notag \\ 
&= \tf f(r_0 )(1-\alpha )x + 2\tf\int_{\rc} ^{r_0}\! \dd r \, \frac{g(r)}{f(r)}
\frac{f(r)^2-f(\rc)f(r_0)}{\sqrt{f(r)^2-f(\rc)^2}}\,, \label{tot}
\end{align}
where we have introduced a dimensionless quantity $\alpha$, 
\begin{equation}
\alpha \equiv \frac{E}{E_{\rm c}}\,, \qquad E_{\rm c} \equiv \tf f(r_0)\,.
\end{equation}
One can show that the total potential vanishes at $\alpha =1$ ($E=E_{\rm c}$) 
and exhibits the critical behavior.
The proof is straightforward. All we have to do is to show that 
\begin{equation}
G(\rc (x))\equiv \int_{\rc} ^{r_0}\! \dd r \, \frac{g(r)}{f(r)}\frac{f(r)^2-f(\rc)f(r_0)}{\sqrt{f(r)^2-f(\rc)^2}}
\end{equation}
is a monotonically decreasing function of $x$ and negative definite. 
Then the second term in (\ref{tot}) 
does not lift up the potential barrier 
and whether the potential barrier vanishes or not 
is relevant only to  the first term in (\ref{tot})\,. 
For $\alpha <1$\,, the potential barrier exists, but it vanishes when $\alpha=1$\,.  
For $\alpha >1$\,, there is no barrier any more. 

\medskip 

It would be helpful to see the critical behavior from numerical plots of the total potential 
$V_{\rm tot}(x)$ for fixed geometries. For $\ads_5\times{\rm S}^5$ and thermal $\ads_5\times{\rm S}^5$\,, 
see Figures 4 and 5 in \cite{SY2}, respectively. For AdS solitons composed of D3-branes and D4-branes, 
see Figures 2 and 3 in \cite{SY3}, respectively. 

\medskip 

Let us then prove that $G(\rc (x) )$ is a monotonically decreasing function of $x$ 
and negative definite. The derivative of $G$ with respect to $x$ is written as  
\[
\frac{\dd G}{\dd x} = \frac{\dd G}{\dd \rc} \cdot \frac{\dd \rc}{\dd x}\,, 
\] 
and each part is given by, respectively, 
\begin{align}
\frac{\dd G}{\dd \rc}&= \left( f(\rc)-f(r_0) \right) \notag \\ 
& \quad \times \lim_{\varepsilon \to 0}
\left( -\frac{g(\rc)}{\sqrt{f(\rc +\varepsilon)^2-f(\rc)^2}}
+\int_{\rc +\varepsilon} ^{r_0}\! \dd r \, \frac{g(r)f(r)f'(\rc)}{(f(r)^2-f(\rc)^2)^{3/2}}\right)\,, \\
\frac{\dd x}{\dd \rc}&=\lim_{\varepsilon \to 0}\left( -\frac{g(\rc)}{\sqrt{f(\rc +\varepsilon)^2-f(\rc)^2}}
+\int_{\rc +\varepsilon} ^{r_0}\! \dd r \, \frac{g(r)f(r)f'(\rc)}{(f(r)^2-f(\rc)^2)^{3/2}}\right)\,.
\end{align}
Here a cutoff parameter $\varepsilon$ has been introduced to regularize the integrals. 
A remarkable point is that the singular parts are canceled out in $\dd G/\dd x$\,. Because $f(r)$ is a monotonically 
increasing function by assumption and $\rc \leq r_0$\,, the following inequality is obtained,  
\begin{equation}
\frac{\dd G}{\dd x}= f(\rc)-f(r_0)\leq0\,.
\end{equation}
Because $G(r_0)=0$\ by definition, where $\rc (x=0)=r_0$\,,  
$G(\rc (x))$ is shown to be a monotonically decreasing function of $x$ and negative definite.

\medskip 

Thus we have shown that $E_{\rm c}$ is the critical value of $E$\,. 
Note that the existence of $E_{\rm c}$ is universal for general backgrounds  
under the conditions supposed here and the value of $E_{\rm c}$ 
depends only on a single function $f(r)$\,.

\subsection{The critical electric field from the DBI action}

Let us here argue the critical electric field from the DBI action for a single D$p$-brane\footnote{
One may consider a higher-dimensional D$q$-brane ($q>p$). Then the extra directions have to be wrapped on 
some non-trivial cycles in the internal manifold. After that, the analysis is similar. }. 
We work in the Lorentzian signature again.  

\medskip 

The NG part of the D$p$-brane action is 
\begin{align}
S_{{\rm D}p} &=-T_{{\rm D}p}\int \! \dd ^{p+1}x\, \sqrt{-\det (g_{\mu \nu}+2\pi \alpha 'F_{\mu \nu} )}\,,  
\end{align}
where $T_{{\rm D}p}$ is the D$p$-brane tension and the world-volume flux $F_{01}=E$ is turned on. 
When the probe D-brane is put at $r=r_0$\,, it can be rewritten as 
\begin{align}
S_{{\rm D}p} &=-T_{{\rm D}p}\int \! \dd ^{p+1}x \, \prod_{i\neq x} G_{ii}(r_0)^{1/2}
\sqrt{ G_{00}(r_0) G_{xx}(r_0) - (2\pi \alpha ')^2E^2 } 
\end{align}
From this expression, one can read off the critical electric field $E_{\rm c}^{\rm (DBI)}$ 
and it agrees with the critical value $E_{\rm c}$ obtained by the potential analysis as follows:  
\begin{equation}
E_{\rm c}^{\rm (DBI)}=\tf \sqrt{G_{00}(r_0)G_{xx}(r_0)}=\tf f(r_0) = E_{\rm c}\,.
\end{equation}
Thus one can see the complete agreement of the critical electric field 
within a class of the gravitational backgrounds under the current consideration (For concrete examples, 
see \cite{SZ,SY2,SY3}). 
This implies a universal behavior of the critical electric field.

\section{Schwinger effects in general confining backgrounds}

We consider here the critical electric field, below which 
the Schwinger effect does not occur, for general confining backgrounds. 
We first give a brief review of 
the sufficient conditions for confining backgrounds argued in \cite{Son1}. 
Then we discuss a universal behavior of the critical electric field under the conditions.

\subsection{Sufficient conditions for confining backgrounds}

Let us first  give a review of the theorem in \cite{Son1}. 
This would be helpful because our later argument heavily depends on this theorem. 
It is useful to rewrite 
\eqref{potential} as 
\begin{eqnarray}
V_{\rm PE+SE} &=& \tf f(\rc)\, x  + 2\tf \int_{\rc}^{r_0}\! \dd r \, \frac{g(r)}{f(r)}\left[ \sqrt{f(r)^2-f(\rc )^2}-f(r)\,\right] 
\nonumber \\
&& \quad - 2\tf \int_{\rt}^{\rc} \! \dd r \, g(r) +2\tf \int_{\rt}^{r_0} \! \dd r \, g(r)\,. 
\end{eqnarray}
The behavior of the potential is essentially determined by the following theorem\footnote{
The original version is slightly modified about the position of the horizon from $r=0$ to $r=\rt$ 
and $\tf$ is also recovered. There is no essential change.}~: 
\vspace*{0.2cm}\\ 
{\bf A theorem to specify confining backgrounds}
\vspace*{0.1cm} 

Let us suppose the following five conditions: 
\begin{enumerate}
\item $f(r)\,,\,g(r) \ge 0$ ~~for~~ $\rt \le r < \infty$\,.
\item $f'(r) > 0$ ~~for~~ $\rt < r < \infty$\,. 
\item $\displaystyle \int^\infty\!\! \dd r \, \frac{g(r)}{f^2(r)}  < \infty$~~i.e. the integral  
converges when its upper limit is sent to infinity. 
\item $f(r)$ is analytic for $\rt  < r < \infty$ and is expanded around $r = \rt$ like 
\begin{equation}
f(r) = f(\rt ) + a_k (r-\rt)^k + \mathcal{O}((r-\rt)^{k+1}) \qquad (k > 0 ~~\&~~  a_k > 0)\,, 
\end{equation}
where $f(r)$ has the minimum at $r=\rt$\,. 
\item $g(r)$ is analytic for $\rt < r < \infty$ and is expanded around $r = \rt$ like 
\begin{equation}
g(r) = b_j (r-\rt)^j + \mathcal{O}((r-\rt)^{j+1}) \qquad  (j > -1 ~~\&~~  b_j > 0)\,.
\end{equation}
\end{enumerate}
Then the potential behavior for sufficiently large $x$ is determined as follows: 
\begin{enumerate}
\item[i)]  If $f(\rt) > 0$ and $k = 2(j+1)$\,, then the linear confinement occurs: 
\begin{equation}
V_{\rm PE+SE} = \tf f(\rt ) x -2 \kappa +2m+\mathcal{O}((\log x)^\beta {\rm e}^{-\gamma x})\,,
\end{equation}
where $\kappa$ is given by 
\begin{equation}
 \kappa \equiv \tf \int_{\rt}^{r_0}\! \dd r \, \frac{g(r)}{f(r)}\left[ f(r)-\sqrt{f(r)^2-f(\rc )^2}\,\right]\,.
\label{kappa}
\end{equation}
This is a positive constant and satisfies the inequality $\kappa < m$\,.
Here $\beta$ and $\gamma$ are positive constants.

\item[ii)]  If $f(\rt) > 0$ and $k > 2(j+1)$\,, then the linear confinement occurs: 
\begin{equation}
V_{\rm PE+SE} = \tf f(\rt ) x -2 \kappa +2m-d x^{-\frac{k+2(j+1)}{k-2(j+1)}}+\mathcal{O}(x^{-\delta})\,,
\end{equation}
where $\kappa$ is the same as \eqref{kappa}, $d$ is a positive constant and $\delta$ is given by
\[
\delta \equiv \frac{k+2(j+1)}{k-2(j+1)}+\frac{2}{k-2j}\,.
\]

\item[iii)] \label{noconf} If $f(\rt) = 0$ and $k > j+1$\,, then the confinement does not occur: 
\begin{equation}
V_{\rm PE+SE}= - A_{\rm s}\, x^{-\frac{j+1}{k-j-1}}+2m+\mathcal{O}(x^{-\delta '})\,,
\end{equation}
where $A_{\rm s}$ is a constant depending on classical string solutions. Here $\delta '$ is given by
\[
\delta ' \equiv \frac{j+1}{k-j-1}+\frac{2k-j-1}{(2k-j)(k-j-1)}\,.
\]
\end{enumerate}
This theorem provides the sufficient conditions to specify general confining backgrounds. 
Note that the first three conditions have already been imposed. Let us suppose the other conditions below  
and we consider general confining backgrounds. 

\subsection*{Examples} 

To illustrate the theorem, let us check the conditions for two examples 
of well-known backgrounds. The first one is AdS solitons, for which the Schwinger effect 
has already been studied in \cite{SY3}. The second is confining backgrounds  
constructed from rotating branes, which have not been discussed in the previous works. 
For more examples, see \cite{Son1}. 

\medskip 

The first example is AdS solitons constructed from D3-branes \cite{HM} . 
The metric in the Lorentzian signature is given by
\begin{align}
\dd s^2=\frac{r^2}{L^2} \left[ -(\dd x^0)^2+\sum _{i=1}^2(\dd x^i )^2+\left( 1-\frac{\rt^4}{r^4}\right)
(\dd x^3 )^2\right]
+\frac{L^2}{r^2}\left( 1-\frac{\rt^4}{r^4}\right)^{-1}\dd r^2+L^2 \dd \Omega _5 ^2\,. \nonumber 
\end{align}
Here the possible range of $r$ is $\rt \leq r < \infty$\,. 
Now suppose that the $x$-direction of the rectangular Wilson loop is taken from either of $x^1$ and $x^2$\,, 
not $x^3$\,. From the metric, the following functions are found, 
\[
G_{00}(r) =G_{xx}(r) = \frac{r^2}{L^2}\,, \quad G_{rr}(r) = \frac{L^2}{r^2}\left(1-\frac{\rt^4}{r^4}\right)^{-1}\,. 
\]
Thus $f(r)$ and $g(r)$ are obtained as 
\begin{eqnarray}
f(r) = \frac{r^2}{L^2}\,, \quad g(r) = \left(1-\frac{\rt^4}{r^4}\right)^{-1/2}\,,
\end{eqnarray}
and  they are manifestly non-negative (The condition 1). 
Then $f(r)$ is monotonically increasing $f'(r) = 2r/L^2 > 0$~(The condition 2). 
It is easy to see the asymptotic behavior, 
\[
\frac{g(r)}{f(r)^2} \sim \frac{L^4}{r^4}~~~(r\to \infty)\,,
\]  
hence the condition 3 is satisfied. 
Note that $f(r)$ has the minimum at $r=\rt$\,. It is expanded around $r=\rt$ like 
\begin{eqnarray}
f(r) = \frac{\rt^2}{L^2} + \frac{2\rt}{L^2}(r-\rt) + \mathcal{O}((r-\rt)^2)\,. 
\end{eqnarray}
Thus $k=1>0$ and $a_1 = 2\rt/L^2 >0$\,, and hence the condition 4 is satisfied. 
Similarly, $g(r)$ is expanded around $r=\rt$ as 
\begin{eqnarray}
g(r) = \frac{\sqrt{\rt}}{2} (r-\rt)^{-1/2} + \mathcal{O}((r-\rt)^{1/2})\,. 
\end{eqnarray}
Hence $j=-1/2> -1$ and $b_{-1/2} = \sqrt{\rt}/2 >0$\,, and the condition 5 is also satisfied. 
Thus the five conditions of the theorem are satisfied. 

\medskip 

It is a turn to check the condition to determine whether confinement occurs or not. 
The minimum of $f(r)$ is positive $f(\rt) = \rt^2/L^2 > 0$\,. 
Because $k=1$ and $j=-1/2$\,, $k=2(j+1)$\,. Thus the AdS soliton backgrounds 
lead to confinement. 

\medskip 

The second example is confining backgrounds constructed from rotating D4-branes 
\cite{Russo,CORT}. These are a generalization of AdS solitons with an additional parameter. 
The metric of this background in the Lorentzian signature is given by
\begin{align}
\dd s^2= &\alpha '\frac{2\pi \lambda A}{3u_0}u\Delta ^{1/2} \left[ 4u^2\eta _{\mu \nu }\dd x^\mu \dd x^\nu
+\frac{4A^2}{9u_0^2}u^2\left( 1-\frac{u_0^6}{u^6\Delta}\right)\dd \theta _2^2 
+ \frac{4}{u^2\left(1-\frac{a^4}{u^4}-\frac{u_0^6}{u^6}\right)}\dd u^2\right. \notag \\
&\left. \qquad \qquad +
 \dd \theta ^2+\frac{\tilde{\Delta}}{\Delta}\sin ^2 \theta \, \dd \varphi ^2+
\frac{1}{\Delta}\cos ^2 \theta \, \dd \Omega _2 ^2-\frac{4a^2Au_0^2}{3u^4\Delta}
\sin ^2 \theta \, \dd \theta _2 \dd \varphi \right]\,. \label{metric2}
\end{align}
Here $\Delta \,, \tilde{\Delta}$ and $A$ are defined as, respectively,  
\begin{equation}
\Delta \equiv 1-\frac{a^4\cos ^2\theta}{u^4}\,, \qquad \tilde{\Delta}\equiv 1-\frac{a^4}{u^4}\,,\qquad
A\equiv \frac{u_0^4}{u_H^4-\frac{1}{3}a^4} \,.
\end{equation}
The parameter $a$ represents the angular momentum and $u_0$ is the horizon when $a=0$\,.
The horizon $u_H$ is determined by the following condition
\begin{equation}
u_H^6-a^4u_H^2-u_0^6=0\,.
\end{equation}
The range of $u$ is $u_H\leq u <\infty$\,.
The metric (\ref{metric2}) leads to the following functions\footnote{
We follow the notation in \cite{Russo,CORT}.
The radial direction is described by $u$\,, not $r$\,. }
\[
G_{00}(u) =G_{xx}(u) = 4C\Delta ^{1/2}u^3\,, \quad 
G_{uu}(u) =\frac{4C\Delta ^{1/2}}{u\left(1-\frac{a^4}{u^4}-\frac{u_0^6}{u^6}\right)}\,, 
\]
where $C$ is a new constant parameter defined as  
\[
C \equiv \alpha '\frac{2\pi \lambda A}{3u_0}\,.
\]
Thus $f(u)$ and $g(u)$ are obtained as 
\begin{eqnarray}
f(u) = 4C\Delta ^{1/2}u^3\,, \quad g(u) = \frac{4C\Delta ^{1/2}u}{\left(1-\frac{a^4}{u^4}-\frac{u_0^6}{u^6}\right)^{1/2}}\,,
\end{eqnarray}
and they are manifestly non-negative (The condition 1). 
Then $f(u)$ is monotonically increasing because 
$f'(u) =u^2\Delta ^{1/2}+2u^2\Delta^{-1/2} > 0$~(The condition 2).   
The conditions 3 is satisfied because of the asymptotic behavior,  
\begin{eqnarray}
\frac{g(u)}{f(u)^2} \sim \frac{1}{4Cu^3} \qquad (u\to \infty)\,. 
\end{eqnarray}
The function $f(u)$ has the minimum at $u=u_H$ and it is expanded around $u=u_H$ as 
\begin{eqnarray}
f(u) =4C\,u_H^3\Delta (u_H)^{1/2} +4C\,u_H^2
\left[
\Delta (u_H)^{1/2}+2\Delta (u_H)^{-1/2}
\right] (u-u_H) + \mathcal{O}((u-u_H)^2)\,. \nonumber 
\end{eqnarray}
Thus $k=1>0$ and $a_1 = 4C\,u_H^2[\Delta (u_H)^{1/2}+2\Delta (u_H)^{-1/2}] > 0$\,, 
and hence the condition 4 is satisfied. 
Similarly, $g(u)$ is expanded around $u=u_H$ as 
\begin{eqnarray}
g(u) = \frac{\Delta (u_H)^{1/2}u_H^4}{\sqrt{6u_H^5-2a^4u_H}} (u-u_H)^{-1/2} + \mathcal{O}((u-u_H)^{1/2})\,. \nonumber
\end{eqnarray}
Hence $j=-1/2> -1$ and $b_{-1/2} = \Delta (u_H)^{1/2}u_H^4/\sqrt{6u_H^5-2a^4u_H} >0$\,, and the condition 5 is also satisfied. 
Thus the five conditions of the theorem are satisfied. 

\medskip 

Then let us check the condition to determine whether confinement occurs or not. 
The minimum of $f(r)$ is positive 
\[
f(u_H) =4Cu_H^2\left[\Delta (u_H)^{1/2}+2\Delta (u_H)^{-1/2} \right] > 0\,.
\] 
Because $k=1$ and $j=-1/2$\,, $k=2(j+1)$\,. Thus the rotating brane backgrounds 
lead to confinement.

\subsection{Critical electric field to liberate quarks \label{3:sec}}

The next task is to consider the Schwinger effect in the case of general confining backgrounds 
specified in the previous subsection.

\medskip 

The Schwinger effect in the confining phase is more intricate than the one in the Coulomb phase, 
due to the existence of the confining potential in addition to the Coulomb potential. It may be 
clear intuitively that the Schwinger effect does not occur when the electric field is weaker than 
the confining string tension.

\medskip 

Then one may expect that the Schwinger effect begins to occur 
above a certain value $E=E_{\rm s}$\,. 
This is indeed the case. The existence of $E=E_{\rm s}$ has been shown explicitly  
by analyzing the total potential in the case of AdS soliton backgrounds 
known as confining backgrounds \cite{SY3}. Note that the existence has been shown for specific examples. 

\medskip 

The result in \cite{SY3} strongly motivates us to show the universal existence of $E_{\rm s}$ 
independently of the detail of the confining backgrounds. To do that, the theorem introduced in the previous subsection 
is quite helpful. In fact, we can show the universal existence and 
the value of $E_{\rm s}$ depends only on a single function $f(r)$\,, similarly to $E_{\rm c}$\,.

\medskip 

Let us consider the confining backgrounds under the conditions. The analysis for arbitrary $x$ seems difficult 
because the confining potential cannot be separated definitely from the other terms. This is a different point from the analysis 
of $E_{\rm c}$\,.  
However, for large $x$\,, it is easy to show that the total potential behaves as
\begin{equation}
V_{\rm tot} ~\sim~ \tf \left[f(\rt ) -\alpha f(r_0) \right] x -2\kappa +2m\,, \qquad \alpha = \frac{E}{E_{\rm c}}\,. 
\end{equation}
This expression leads to the critical electric field, 
\begin{eqnarray}
E= \tf f(\rt) \equiv E_{\rm s}\,. 
\end{eqnarray}
Note that $E_{\rm s}$ depends only on $f(r)$ as in the case of $E_{\rm c}$\,.

\medskip 

The condition that $f(r)$ is a monotonically increasing function gives rise to 
the inequality 
\[
E_{\rm s} < E_{\rm c}\,. 
\]
Therefore $E_{\rm s}$ does not coincide with $E_{\rm c}$\,. 
It is obvious that this inequality comes from the geometric relation $\rt < r_0$\,.

\section{Conclusion and Discussion}

We have performed the potential analysis for studying the holographic Schwinger effect 
in general backgrounds and have shown universal behaviors for two kinds of the critical electric field. 

\medskip 

The first is the critical electric field $E_{\rm c}$, above which the potential barrier vanishes 
and the system becomes unstable catastrophically. The existence of it is universal 
for a wide class of the gravitational backgrounds under some assumptions. 
The critical electric field obtained from the potential analysis  
agrees with the one obtained from the DBI action. 

\medskip 

The second is the critical electric field $E_{\rm s}$\,, below which 
the Schwinger effect does not occur any more. It universally exists for general confining backgrounds  
satisfying the sufficient conditions presented in \cite{Son1}. 
It is worth noting that the behavior of this type is supported by a recent numerical simulation 
in lattice QCD \cite{lattice}.

\medskip 

So far, the confining backgrounds have been considered under the sufficient conditions. 
It is interesting to consider some exceptional cases,  
where all of the conditions are not satisfied but confinement occurs. 
For example, TsT transformations change the boundary behavior of the bulk spacetime 
and hence it may break some of the conditions.   
It is also nice to consider a holographic argument 
for non-abelian Schwinger effects \cite{na1,na2,na3}. 

\medskip 

Another interesting issue is to study a connection between the Schwinger effect and 
other non-perturbative phenomena like chiral symmetry breaking 
by considering more QCD-like setups such as 
the Sakai-Sugimoto model \cite{SS} and D3/D7-brane systems\footnote{
For some discussions on a D3/D7-brane system, see \cite{Hashi}.}.

\section*{Acknowledgments}

We would like to thank Y.~Hidaka for useful discussions. 
This work was also supported in part by the Grant-in-Aid 
for the Global COE Program ``The Next Generation of Physics, Spun 
from Universality and Emergence'' from MEXT, Japan.

\end{document}